\documentclass[acmtog, screen, nonacm]{acmart}

\acmSubmissionID{608}

\usepackage{xcolor}
\usepackage{float}
\usepackage{bm}
\usepackage{cleveref}
\usepackage{acronym}
\usepackage{wrapfig}

\setcopyright{none}
\setcitestyle{square}

\copyrightyear{2024}
\acmYear{2024}
\setcopyright{acmlicensed}
\acmConference[SIGGRAPH Conference Papers '24]{Special Interest Group on Computer Graphics and Interactive Techniques Conference Conference Papers '24}{July 27-August 1, 2024}{Denver, CO, USA}
\acmBooktitle{Special Interest Group on Computer Graphics and Interactive Techniques Conference Conference Papers '24 (SIGGRAPH Conference Papers '24), July 27-August 1, 2024, Denver, CO, USA}
\acmDOI{10.1145/3641519.3657448}
\acmISBN{979-8-4007-0525-0/24/07}

\begin{document}

\title{VR-GS: A Physical Dynamics-Aware Interactive Gaussian Splatting System in Virtual Reality}

\author{Ying Jiang}
\orcid{0000-0002-7026-0073}
\authornote{\ indicates equal contributions.}
\affiliation{
\institution{UCLA}
\country{USA}
}
\affiliation{
\institution{HKU}
\country{Hong Kong}
}
\email{anajymua@gmail.com}

\author{Chang Yu}
\orcid{0009-0000-2613-9885}
\authornotemark[1]
\affiliation{
\institution{UCLA}
\country{USA}
}
\email{g1n0st@live.com}

\author{Tianyi Xie}
\orcid{0009-0006-3101-7659}
\authornotemark[1]
\affiliation{
\institution{UCLA}
\country{USA}
}
\email{tianyixie77@g.ucla.edu}

\author{Xuan Li}
\orcid{0000-0003-0677-8369}
\authornotemark[1]
\affiliation{
\institution{UCLA}
\country{USA}
}
\email{xuanli1@g.ucla.edu}

\author{Yutao Feng}
\orcid{0009-0003-3700-8102}
\affiliation{
\institution{University of Utah}
\country{USA}
}
\affiliation{
\institution{Zhejiang University}
\country{China}
}
\email{fytal0n@gmail.com}

\author{Huamin Wang}
\orcid{0000-0002-8153-2337}
\affiliation{
\institution{Style3D Research}
\country{China}
}
\email{wanghmin@gmail.com}

\author{Minchen Li}
\orcid{0000-0001-9868-7311}
\affiliation{
\institution{CMU}
\country{USA}
}
\email{minchernl@gmail.com}

\author{Henry Lau}
\orcid{0000-0002-0786-9249}
\affiliation{
\institution{HKU}
\country{Hong Kong}
}
\email{hyklau@hku.hk}

\author{Feng Gao}
\orcid{0000-0003-1515-1357}
\authornote{This work is not related to F. Gao's position at Amazon.}
\affiliation{
\institution{Amazon}
\country{USA}
}
\email{fenggo@amazon.com}

\author{Yin Yang}
\orcid{0000-0001-7645-5931}
\affiliation{
\institution{University of Utah}
\country{USA}
}
\email{yangzzzy@gmail.com}

\author{Chenfanfu Jiang}
\orcid{0000-0003-3506-0583}
\affiliation{
\institution{UCLA}
\country{USA}
}
\email{cffjiang@ucla.edu}

\renewcommand{\shortauthors}{Y. Jiang, C. Yu, T. Xie, X. Li, Y. Feng, H. Wang, M. Li, H. Lau, F. Gao, Y. Yang, C. Jiang}
\acrodef{cv}[CV]{Computer Vision}
\acrodef{cg}[CG]{Computer Graphics}
\acrodef{aigc}[AIGC]{Artificial Intelligence Generated Content}
\acrodef{vr}[VR]{Virtual Reality}
\acrodef{mr}[MR]{Mixed Reality}
\acrodef{nerf}[NeRF]{Neural Radiance Fields}
\acrodef{gs}[GS]{Gaussian Splatting}
\acrodef{pde}[PDE]{Partial Differentiable Equation}
\acrodef{pbd}[XPBD]{eXtended Position-based Dynamics}

\begin{abstract}
As 3D content becomes increasingly prevalent, there's a growing focus on the development of engagements with 3D virtual content. Unfortunately, traditional techniques for creating, editing, and interacting with this content are fraught with difficulties. They tend to be not only engineering-intensive but also require extensive expertise, which adds to the frustration and inefficiency in virtual object manipulation. 
Our proposed VR-GS system represents a leap forward in human-centered 3D content interaction, offering a seamless and intuitive user experience. By developing a physical dynamics-aware interactive \acf{gs} in a \ac{vr} setting, and constructing a highly efficient two-level embedding strategy alongside deformable body simulations, VR-GS ensures real-time execution with highly realistic dynamic responses.
The components of our system are designed for high efficiency and effectiveness, starting from detailed scene reconstruction and object segmentation, advancing through multi-view image in-painting, and extending to interactive physics-based editing. The system also incorporates real-time deformation embedding and dynamic shadow casting, ensuring a comprehensive and engaging virtual experience.
\end{abstract}



\begin{teaserfigure}
  \includegraphics[width=\textwidth]{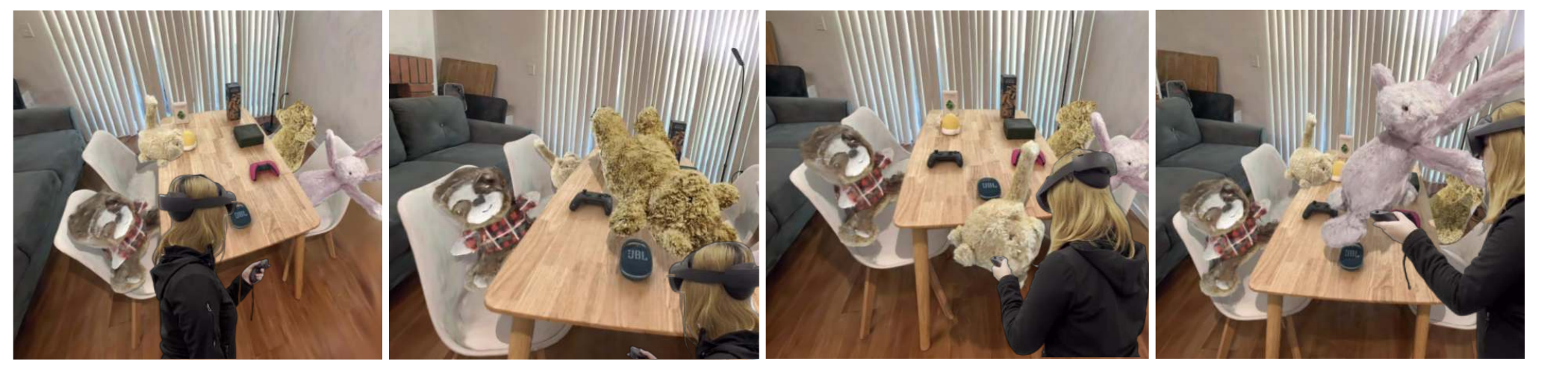}
  \caption{\textbf{Animal Crossing.} Utilizing our system, individuals can engage in intuitive and interactive physics-based game-play with deformable virtual animals and realistic environments represented with 3D Gaussian Splatting.}
  \label{fig:teaser}
\end{teaserfigure}

\maketitle

\section{Introduction} \label{sec:intro}

As digital technology advances, the importance of 3D content is becoming increasingly prominent across various industries, from entertainment to education. This growth is driving the demand for high-fidelity 3D content within the \ac{cg} and \ac{cv} communities. Despite their capabilities of being rendered efficiently, traditional 3D/4D content creation, reliant on 3D modeling tools and game engines, is time-intensive and complex, often beyond the reach of non-expert users. This accessibility barrier limits broader user engagement in high-quality content creation.

Recognizing the advantages and limitations of the traditional graphics pipeline, our goal is to find a modern variant. To enhance the visual quality and ease of creation for non-expert users, we move away from traditional 3D models in graphics pipelines and instead, adopt state-of-the-art radiance field techniques for rendering. In this context, \acf{nerf} emerge as a natural choice. Despite its versatility, \ac{nerf}'s volume rendering falls short in efficiency for interactive applications, which demand high frame rates. Additionally, \ac{nerf}'s approach to handling deformations—requiring bending of query rays via an inverse deformation map—is slow \cite{pumarola2021d}. Fortunately, 3D \acf{gs} \cite{kerbl20233d} has been recently introduced as an efficient and explicit alternative to \ac{nerf}. This method not only excels in rendering efficiency but also provides an explicit geometric representation that can be directly deformed or edited. The explicit nature of 3D \ac{gs} can simplify the direct manipulation of geometry by solving a \acf{pde} that governs the motions of Gaussian kernels. Furthermore, \ac{gs} eliminates the need for high-fidelity meshes, UV maps, and textures, offering the potential for naturally photorealistic appearances. It is worth noting that many studies have already demonstrated the use of 3D \ac{gs} for 4D dynamics \cite{park2021nerfies, pumarola2021d, yang2023real} and the integration of physics-based simulations for more realistic 4D content generation \cite{xie2023physgaussian, feng2023pie}, as well as for creating animatable avatars \cite{zielonka2023drivable, xu2023gaussian}.

In this paper, we introduce a physics-aware interactive system for immersive manipulation of 3D content represented with 
\ac{gs}. To ensure an interactive experience, we utilize \acf{pbd} \cite{macklin2016xpbd}, a highly adaptable and unified physical simulator, for real-time deformation simulation. Direct incorporation of a simulator onto \ac{gs} kernels presents challenges, as the simulation and rendering processes have distinct geometrical representations. To address this, we construct a tetrahedral cage for each segmented \ac{gs} kernel group and embed these kernel groups into corresponding meshes. The deformed mesh, driven by \ac{pbd}, subsequently guides the deformation of the \ac{gs} kernels. Noticing that simplistic embedding techniques can lead to undesirable, spiky deformations in \ac{gs} kernels, we propose a novel two-level embedding approach. This method allows each Gaussian kernel to adapt to a smoothed average deformation of the surrounding tetrahedra. The intricate combination of \ac{gs} and \ac{pbd} through our two-level embedding not only achieves real-time physics-based dynamics but also upholds high-quality, realistic rendering. In summary, our contributions include:
\begin{itemize}
\item \emph{A Physics-Aware Interactive System:} Development of a system that enables interactive, physics-aware manipulation of 3D content represented with \ac{gs}.
\item \emph{Two-Level Deformation Embedding:} Introduction of a novel two-level embedding approach that allows Gaussians to adapt smoothly to the mesh, enhancing the deformation realism and preventing undesirable spiky artifacts.
\end{itemize}
We deploy our system on VR devices, enriched by segmentation, inpainting, and shadow map, offering users a rich platform for 3D content manipulation.

\section{Related Work}\label{sec:rw}

\subsection{Radiance Fields Rendering} 
A variety of 3D representations, such as mesh \cite{sorkine2004least}, point clouds \cite{qi2017pointnet}, signed distance fields \cite{wang2021neus}, and grids \cite{zhu2005animating} are exploited in early work for visual computing tasks, including synthesis, estimation, manipulation, animation, reconstruction, and transmission of data about objects and scenes \cite{kerbl20233d, xie2022neural, gao2023scenehgn}. 
Since neural radiance fields (NeRF) \cite{mildenhall2021nerf} was proposed for novel view synthesis with differentiable volume rendering, it has been applied to versatile computer graphics and computer vision tasks, such as scene reconstruction \cite{oechsle2021unisurf, meng2023neat, yariv2021volume}, synthesis \cite{sun2023sol, huang2023nerf}, rendering \cite{lombardi2021mixture, wu2023nerf}, interactive games \cite{xia2024video2game} and animation \cite{chen2021animatable, peng2021animatable}, simulation \cite{li2023pac}, etc. While Neural Radiance Fields (NeRF) have achieved remarkable image quality, they suffer from significant time and memory inefficiencies \cite{lindell2021autoint}. To mitigate these issues, various methods have been introduced.  Sparse representations \cite{liu2020neural}, decomposed strategies \cite{niemeyer2021campari, rebain2021derf}, and multi-resolution encoding \cite{muller2022instant} have all been proposed to enhance volume rendering efficiency while maintaining high quality. However, NeRFs are implicit representations, making it challenging to detect and resolve collisions \cite{qiao2023dynamic}. Recently, 3D Gaussian splatting (GS) proposed by \cite{kerbl20233d} utilizes an array of 3D Gaussian kernels to represent scenes explicitly, facilitating faster optimization and achieving state-of-the-art results.

\subsection{Editing in Radiance-based Scene}
A natural extension of NeRF is to support user-guided editing. \citet{li2023interactive} made use of two proxy cages to offer interactive control of the shape deformation of NeRF. Besides geometry editing, text prompts \cite{wang2023nerf} or a user-edited image \cite{bao2023sine} can also be adopted to conduct style transfer of NeRF. Besides, \citet{lin2023sketchfacenerf} put forward exploiting 2D sketches to edit high-quality facial NeRF. Concerning NeRF texture editing, \citet{huang2023nerf} utilized a coarse-fine disentanglement representation and a patch-matching algorithm to synthesize textures of different shapes from multi-view images. De-Nerf \cite{wu2023nerf} exploited a hybrid light representation that enables users to relight NeRF. In contrast, Gaussian Splatting is more appropriate for post-editing tasks thanks to its explicit nature and has inspired a series of follow-up works \cite{chen2023gaussianeditor, fang2023gaussianeditor, ye2023gaussian, duisterhof2023md, huang2023sc}.

Another significant direction in NeRF research is the incorporation of dynamic components into static scenes. To achieve this, a popular strategy \cite{pumarola2021d, park2021nerfies} is to consider time as an additional input to the system, This method typically splits a dynamic NeRF into a canonical static field and a deformation field, with the latter mapping this canonical representation to a deformed one. More recently, \citet{yang2023real} drew inspiration from 3D GS to reconstruct a dynamic scene using 4D Gaussian primitives that change over time. However, the dynamics in these methods are generally confined to motions captured from input data, limiting their ability to synthesize unseen dynamics. To generate novel dynamics, \citet{xie2023physgaussian} and \citet{feng2023pie} have integrated physics-based simulations into 3D static representations, paving the way for generating physically plausible and novel dynamic scenes.

\subsection{Real-time Neural Radiance Fields}
Rendering NeRF is computationally expensive for real-time applications, such as VR, which causes high latency and low quality \cite{song2023nerfplayer, li2022immersive}. To address these challenges and enable high-fidelity rendering with minimal latency on a single GPU, various techniques have been proposed. These include gaze-contingent 3D neural representations \cite{deng2022fov}, variable rate shading \cite{rolff2023vrs}, and hybrid surface-volume representations \cite{turki2023hybridnerf}, all aimed at accelerating NeRF. In contrast to single-GPU solutions, VR-NeRF \cite{xu2023vr} leveraged multiple GPUs to achieve high-quality volumetric rendering of NeRF. Beyond software acceleration techniques, RT-NeRF \cite{li2022rt} utilized an algorithm-hardware co-design framework to provide real-time NeRF solutions for immersive VR rendering. While these methods primarily focus on enhancing NeRF rendering, our work is dedicated to facilitating interactive editing.

\citet{li2023interacting} integrated affine transformation with NeRF to enable interactive features, such as exocentric manipulation, editing, and VR tunneling effects.  However, their approach is limited to filter-based edits such as edge blurring and fails to support the deformation of virtual objects' geometry. RealityGit \cite{li2023realitygit} and Magic NeRF Lens \cite{li2023magic} modified a NeRF model by erasing or revealing a portion of NeRF. Nevertheless, neither system permits deforming the geometrical structure of the model. In contrast to these works which concentrate on interactive editing of NeRF, our system enables users to interact with 3D GS in a physically realistic way in real time.

\section{System Design}\label{sec:design}

We propose a unified physics dynamics-aware interactive \ac{vr} system for real-time interactions with \acf{gs}.

\begin{figure*} [ht]
 \includegraphics[width=1.0\textwidth]{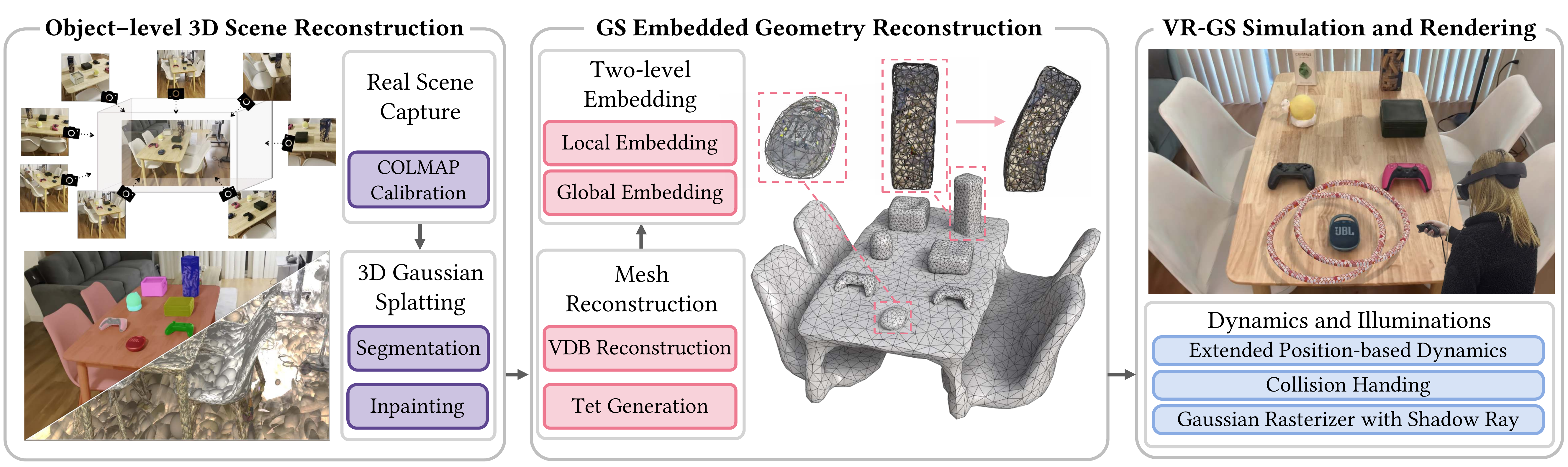}
  \centering
  \caption{
  VR-GS is an interactive system designed to integrate 3D \acf{gs} and \acf{pbd} for generating a real-time interactive experience. 
  Beginning with multi-view images, the pipeline combines scene reconstruction, segmentation, and inpainting using Gaussian kernels. These kernels form the foundation for VR-GS's utilization of the sparse volumetric data structure VDB, facilitating bounding mesh reconstruction and subsequent tetrahedralization. VR-GS further harnesses a novel two-level Gaussian embedding, \ac{pbd}, collision detection, and shadow casting techniques, all converging to deliver a captivating and immersive user experience.
  }
  \label{fig:pipeline}
\end{figure*}

\paragraph{Immersive and Realistic Generative Dynamics}
Our targeted system aims to offer users an immersive and realistic experience, characterized by dynamic motions, 3D virtual shapes, and illuminations that closely mirror the real world \cite{kalawsky1999vruse}. This goal sets our system apart, especially in how it handles physics-based 3D deformations and dynamics, as opposed to those that rely solely on geometric deformation. The proposed system leverages physics-based simulation techniques to produce realistic dynamics and facilitate 3D shape editing. Unlike methods that reconstruct motion from time-dependent datasets or utilize generative AI to drive Gaussian models, our system adheres to physical principles. To create an immersive experience, we choose VR as the interaction platform.

\paragraph{Real-Time Interaction}
For an interactive system, low latency is crucial to prevent disorientation or motion sickness \cite{sutcliffe2019reflecting, deng2022fov}. To offer instantaneous output and feedback, most interactive systems support basic transformations including rotation, scaling, and translation. Our  system additionally offers users real-time physics-based interaction. To our knowledge, VR-GS represents the first interactive physics-based \ac{gs} editor. To achieve both real-time performance and physically realistic editing, we implement a reduced representation model and parallel-friendly algorithms within our frame budget. While the per-Gaussian-based discretization as done in PhysGaussian \cite{xie2023physgaussian} offers detailed dynamics, its time cost is impractical for our system's needs. Instead, we utilize a tetrahedral mesh embedding reconstructed from Gaussian kernels to drive \ac{gs} motion. We adopt \ac{pbd} with finite-element constraints to achieve real-time simulations.

\paragraph{Unified Framework} 
VR-GS is a framework that integrates rendering and simulation within a unified pipeline. Our design principles adhere to the goal of ``what you see is what you simulate'' \cite{SimulatingVisualGeometry}. Unlike \citet{xie2023physgaussian}, our system employs a hybrid approach, combining cage mesh and Gaussian kernels, to achieve real-time performance. The deformation gradient field of Gaussian kernels is embedded in the cage mesh through piecewise constant interpolation. With each Gaussian kernel resembling a rotated ellipsoid, this simple embedding strategy leads to spiky artifacts during large deformation, a challenge highlighted in \cite{xie2023physgaussian}. To mitigate this, we employ a two-level interpolation scheme: initially embedding each Gaussian within an individual bounding tetrahedron, followed by embedding these tetrahedra within the simulated cage mesh to elevate the smoothness of the deformation field.

\paragraph{Additional Components}
Our framework is designed to construct VR-compatible interactive settings derived from real-world captures. Consequently, it introduces complexities associated with facilitating intuitive user interactions and delivering real-time feedback that meets user expectations and common sense perceptions \cite{stanney2003usability, hale2014handbook, gabbard1999user}. In particular, when users engage with objects within a scene, they anticipate natural movements and realistic appearances from both the selected objects and the surrounding environment. Moreover, it is necessary to address voids when an object is removed from its supporting base. To manage these challenges, our system additionally integrates functionalities for object segmentation and scene inpainting.

\section{Method}\label{sec:method}
\subsection{Gaussian Splatting}
Gaussian splatting, proposed by \citet{kerbl20233d}, is an explicit 3D representation to encapsulate 3D scene information using a set of 3D anisotropic Gaussian kernels, each with learnable mean $\mu$, opacity $\sigma$, covariance matrix $\Sigma$, and spherical harmonic coefficients $\mathcal{C}$. Spherical harmonics (SH) are expansions of the view-dependent color function defined on the unit sphere. 
To render a view, the 3D splats are projected to 2D screen space ordered by their $z$-depth. The color $C$ of a pixel is computed by $\alpha$-blending of these 2D Gaussians from near to far:
\begin{equation}\label{euqation:gs}
    C = \sum_{i \in {N}} c_{i}\alpha_{i}\prod_{j=1}^{i-1} (1 - \alpha_{j}),
\end{equation}
where $c_{i}$  is the evaluated color by SHs viewed from the camera to the kernel's mean. $\alpha_{i}$ is the product of the kernel's opacity and 2D Gaussian weight evaluated at the pixel coordinate. Leveraging a differentiable implementation, the rendering loss towards the ground truth image can be backpropagated to Gaussians' parameters for optimization.

In contrast to traditional NeRFs based on implicit scene representations, \ac{gs} provides an explicit representation that can be seamlessly integrated with post-processing manipulations, such as animating and editing. Moreover, the efficient rasterization and superior rendering quality of 3D Gaussians facilitate their integration with VR.

\subsection{VR-GS Assets Preparation}
Each 3D asset in our \ac{vr} system consists of a high-fidelity 3D \ac{gs} reconstruction and an envoloping simulatable tetrahedral mesh in a moderate resolution to enable real-time physics-aware dynamics. These preparations are conducted offline before immersive and interactive editings within our \ac{vr} environment, which includes:
\begin{itemize}
    \item \emph{Segmented \ac{gs} Generation}: we support interactions with individual objects in a large scene, achieved by segmented \ac{gs} reconstructions.
    \item \emph{Inpainting}: The occluded parts between the objects and their supporting planes usually have no texture. We inpaint 3D \ac{gs} representations leveraging a 2D inpainting technique.
    \item \emph{Mesh Generation}: A simulation-ready tetrahedral mesh is generated for each object.
\end{itemize}

\subsubsection{Segmentation}
The segmented \ac{gs} is constructed during \ac{gs} training. We first generate 2D masks on the multi-view RGB images by utilizing a 2D segmentation model \cite{cheng2023putting}. Each segmented part is assigned a different color that is consistent across different views. Subsequently, we enhance the scene representation by integrating three additional learnable RGB attributes into the 3D Gaussian kernels. During the reconstruction process, each 3D Gaussian kernel will automatically learn what object it belongs to  utilizing a segmentation loss function $L_{\text{seg}}$:
\begin{equation}
    L_{\text{seg}} = L_{1}(M_{2d}, I),
\end{equation}
where $M_{2d}$ represents colored 2D segmentation results and $I$ denotes the rendering of 3D Gaussian kernels with colors replaced by the extended RGB attributes instead of evaluated from SHs. Thus, the total loss becomes
\begin{equation}
    L_{\text{total}} =  (1- \lambda)L_{1} + \lambda L_{\text{SSIM}} + \lambda_{\text{seg}} L_{\text{seg}},
\end{equation}
where $L_{1}$ and $L_{\text{SSIM}}$ are computed between the normally rendered images and the multi-view ground truths.
We use $\lambda = 0.2$ and $\lambda_{\text{seg}} = 0.1$ in all our experiments.

\subsubsection{Inpainting}
After 3D GS segmentation, we extract all objects separately from the scene. This process of object removal, however, results in the emergence of holes within the regions that are previously occluded. To alleviate the issue, we utilize a 2D inpainting tool LaMa \cite{suvorov2022resolution} to guide the 3D inpainting of Gaussian kernels. We freeze the Gaussian kernels located outside of the holes and then use an inpainting loss $L_{\text{inpaint}}  = L_1(I_{\text{inpainted}}, I)$ to optimize a  Gaussian kernel patch under the guidance of the 2D inpainted images $I_{\text{inpainted}}$ for the current 3D \ac{gs} rendering $I$. The result of our inpainting is validated in \Cref{sec:ablation_shadow_map}.

\subsubsection{Mesh Generation} 
Due to our design choice to use mesh-based simulation, we generate a tetrahedral mesh for each group of the segmented 3D GS kernels. These meshes will not be rendered during the interaction but only serve as the media of dynamics. Hence we can use a moderate mesh resolution which does not hinder performance. To construct a simulation mesh, we first use internal filling proposed by \cite{xie2023physgaussian} to fill particles into the void internal region that is not reconstructed by \ac{gs}. Then we treat centers of Gaussians as a point cloud and convert it to a voxelized VDB representation \cite{museth2013vdb}. A water-tight surface mesh is then extracted using marching cubes \cite{lorensen1987marching} and tetrahedralized into a finite element mesh using TetGen \citep{tetgen}. 

\begin{figure}
  \includegraphics[width=1.0\linewidth]{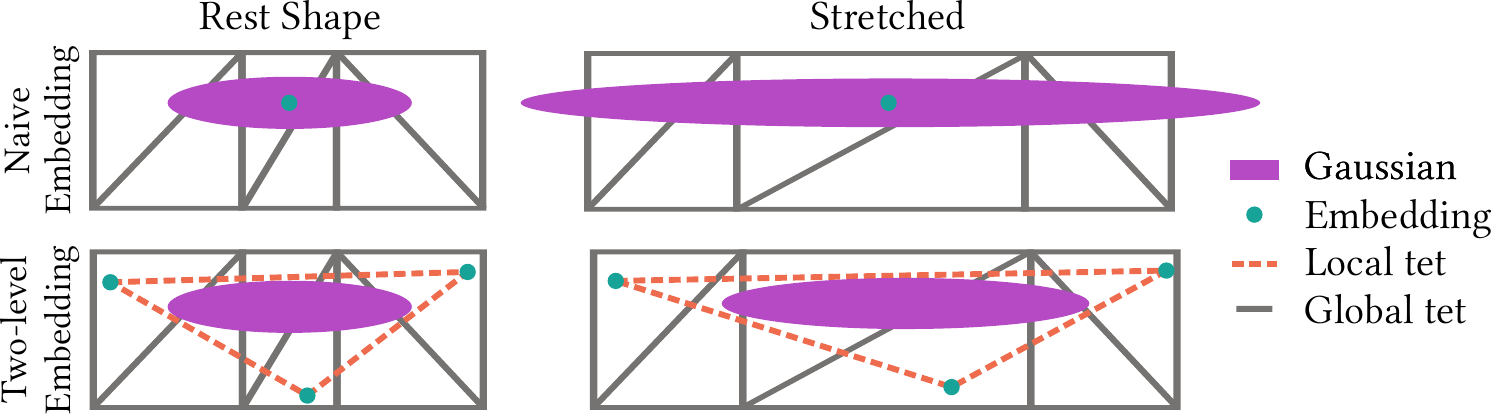}
  \caption{Our two-level embedding effectively resolves spiky artifacts. Each Gaussian kernel is embedded into a local tetrahedron. The vertices of the local tetrahedron are independently embedded into the global mesh.}
  \label{fig:spike}
\end{figure}

\subsection{Unified Framework for Simulation and Rendering}

As shown in \citet{xie2023physgaussian}, Gaussian kernels can be deformed by the simulated deformation field. We follow the Gaussian kinematics of this work but replace the simulator with XPBD \cite{macklin2016xpbd} to achieve real-time interactions. We employ the strain energy constraint as the elastic model and adopt the velocity-based damping model in the XPBD framework. 

\subsubsection{Physical Parameters} \label{sec:physparam}
The physical parameters, such as densities and material stiffness, are tunable hyperparameters. Following PhysGaussian \cite{xie2023physgaussian}, we manually set physical parameters for all interactable objects, including Young's modules ($E$), Poisson ratio ($\nu$), and density ($\rho$). We start from an initial configuration $E=1000\text{ Pa}, \nu=0.3, \rho=1000 \text{ kg}$ for each object, then tune them to produce visually plausible dynamics. The Young's modulus governs the overall stiffness of an object and is manually adjusted by analyzing the severity of deformation of its free fall collision with the ground. Additionally, we adjust the relative density between two objects based on their deformations under collision. In practice we usually arrive at a finalized parameter setting that gives visually plausible results within 10 iterations of this manual process.

\subsubsection{Embedding}
In our mesh-based simulation, the deformation map is piece-wise linear, with the resulting deformation gradient piece-wise constant within each tetrahedron. Given a tetrahedron with the rest-shape configuration $\{\bm x_0^0, \bm x_1^0, \bm x_2^0, \bm x_3^0\}$ and the current configuration $\{\bm x_0, \bm x_1, \bm x_2, \bm x_3\}$, the deformation gradient is defined as 
\begin{equation}
    \bm F = \begin{bmatrix}\bm x_1 - \bm x_0, \bm x_2 - \bm x_0 , \bm x_3 - \bm x_0\end{bmatrix} \begin{bmatrix}\bm x_1^0 - \bm x_0^0 , \bm x_2^0 - \bm x_0^0 , \bm x_3^0 - \bm x_0^0\end{bmatrix}^{-1},
\end{equation}
where the inverse of the rest-shape basis can be computed pre-simulation. The mean and the covariance matrix of the deformed Gaussian kernel inside this tetrahedron is given by
\begin{equation}
    \bm \mu = \sum_{i} w_i \bm x_i, \quad \bm \Sigma = \bm F \bm \Sigma_0 \bm F^T, 
\end{equation}
where $ w_i$ is the barycentric coordinates of the initial center $\bm \mu_0$ in the rest-shape configuration, and  $\bm \Sigma_0$ is the initial covariance matrix \cite{xie2023physgaussian}.  The deformed Gaussian kernels can be directly rendered by the point splatting procedure. However, the direct embedding of Gaussian centers cannot guarantee that every ellipse shape is completely enveloped by some tetrahedron inside the simulation mesh. As shown in \Cref{fig:spike}, this can lead to spiky artifacts. Observe that kernels that are completely inside the tetrahedron will always be inside. This motivates us to propose a two-level embedding procedure:
\begin{enumerate}
    \item Local embedding: we independently envelope every Gaussian kernel by an as-tight-as-possible tetrahedron.  There is no connectivity between these local tetrahedra.
    \item Global embedding: we embed the vertices of local tetrahedra into the global simulation mesh. 
\end{enumerate}
As the global mesh is deformed by the simulation, the vertices of local tetrahedra are kept inside the boundary, driving the kinetic evolution of Gaussian kernels. A local tetrahedron could overlap with multiple global tetrahedra. The deformation map on it can be understood as the average of the surrounding global tetrahedra, hence eliminating sharp, spiky artifacts, as validated in Section \ref{sec:ablation_embedding}.

\subsubsection{Shadow Map} While the original global illumination of the scene can be accurately learned and baked by the spherical harmonics on each Gaussian, the shadow will no longer be aligned with the object when it is moving or deforming. Bringing shadow map \cite{shadowmap} into the GS framework can enhance the immersive experience in \ac{vr}. More importantly, it can guide human perception of the spatial relationships between objects: during manipulation, users will rely on the shadow to determine the distance between objects. The shadow map is a fast real-time algorithm and is well-aligned with the GS rasterization pipeline. We follow \Cref{euqation:gs} to estimate the depth map from the light source and test the visibility for each Gaussian using this depth map. The influence of shadow maps is studied in \Cref{sec:ablation_shadow_map}.

\begin{figure}
  \includegraphics[width=1.0\linewidth]{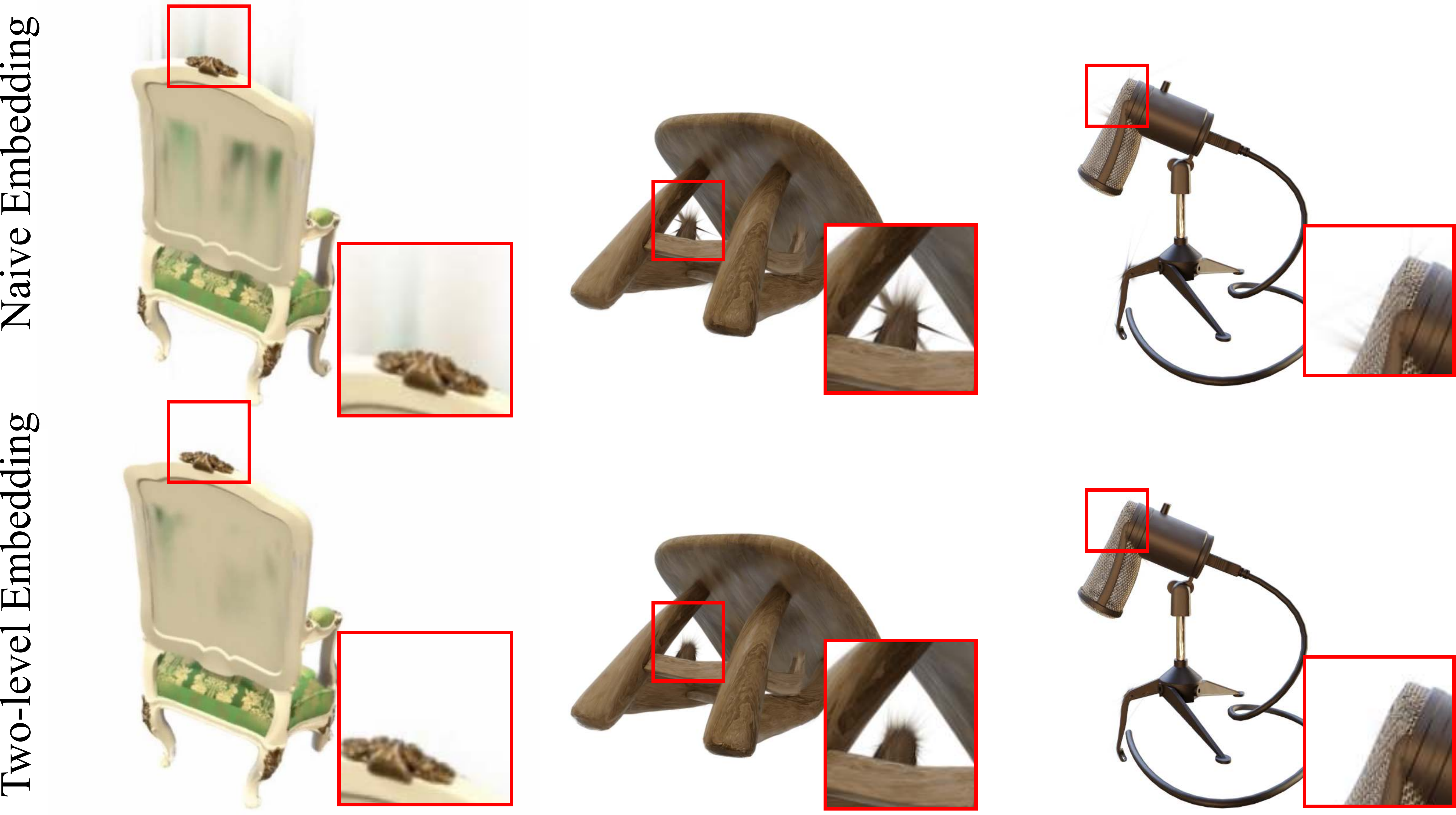}
  \caption{\textbf{Two-level Embedding Evaluation.} Our two-level embedding alleviates the commonly seen spiky artifacts for deformed GS kernels.}
  \label{fig:two-level embedding}
\end{figure}

\begin{figure}
  \includegraphics[width=1.0\linewidth]{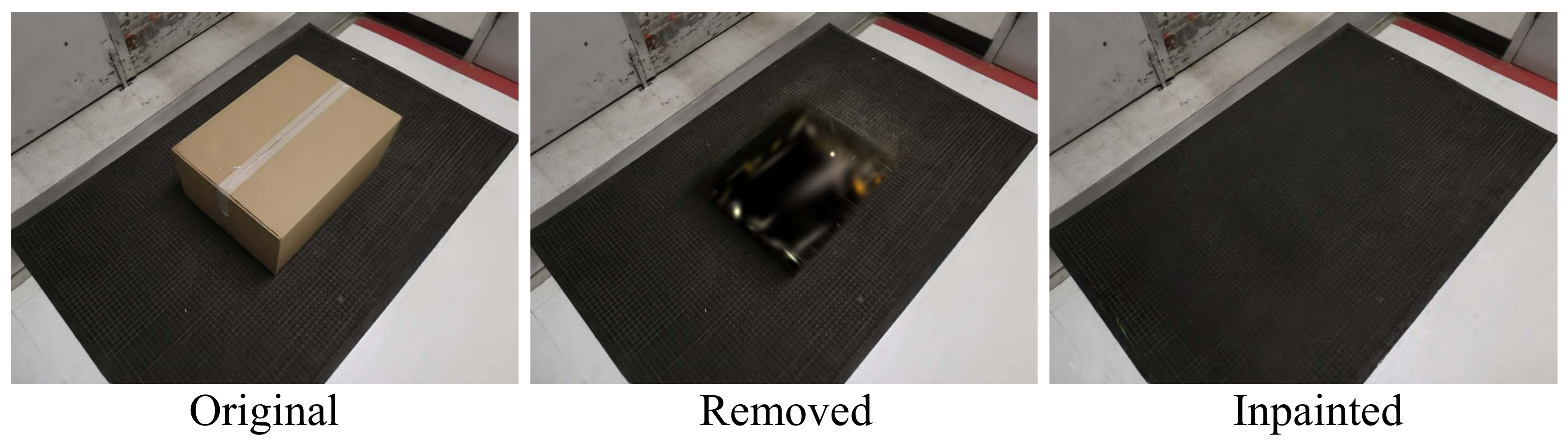}
  \caption{\textbf{Inpainting Evaluation.}
  \ac{gs} struggles to reconstruct occluded surfaces. By leveraging LAMA \cite{suvorov2022resolution}, we produce 2D inpainted results that guide the 3D scene inpainting, enhancing the realism of the 3D representation.}
  \label{fig:inpainting}
\end{figure}

\section{Evaluation}\label{sec:eval}
In this section, we assess the essential components of our proposed system, including two-level \ac{gs} embedding and shadow map. 

\paragraph{Two-level Embedding}\label{sec:ablation_embedding}
The two-level embedding is a crucial component in our physical dynamics-aware system, integrating the tetrahedra cage mesh with the embedded Gaussian. In \Cref{fig:two-level embedding}, we conduct an ablation study to validate the effectiveness of our two-level embedding approach. Under conditions of extreme stretching or twisting, the naive embedding method, which simply embeds the Gaussian kernel within the closest tetrahedron, leads to severe spiky artifacts. Our two-level embedding strategy addresses this by initially embedding each Gaussian into a localized, independent tetrahedron, followed by embedding the vertices of this tetrahedron into the cage mesh. The deformation of each Gaussian kernel is determined by averaging the deformations at these vertices, resulting in a smoother deformation gradient field than the naive approach and significantly reducing spiky artifacts.

\paragraph{Inpainting}\label{sec:ablation_inpainting}
With the guidance of 2D image segmentation results, our system achieves object-level 3D scene reconstruction, facilitating convenient post-physics-based manipulation. However, GS is limited to reconstructing surfaces visible in the provided multi-view training images. Consequently, some object and background areas, unseen in these images, remain unreconstructed by GS, leading to ``black hole'' like artifacts when foreground objects are moved. To address this, our system integrates the object segmentation mask with LAMA \cite{suvorov2022resolution}, a 2D inpainting model, to generate inpainted multi-view images. These images then guide the fine-tuning and inpainting of our 3D GS scene, yielding a more complete and realistic user experience, as evidenced in Figure \ref{fig:inpainting}.

\begin{figure}
  \includegraphics[width=1.0\linewidth]{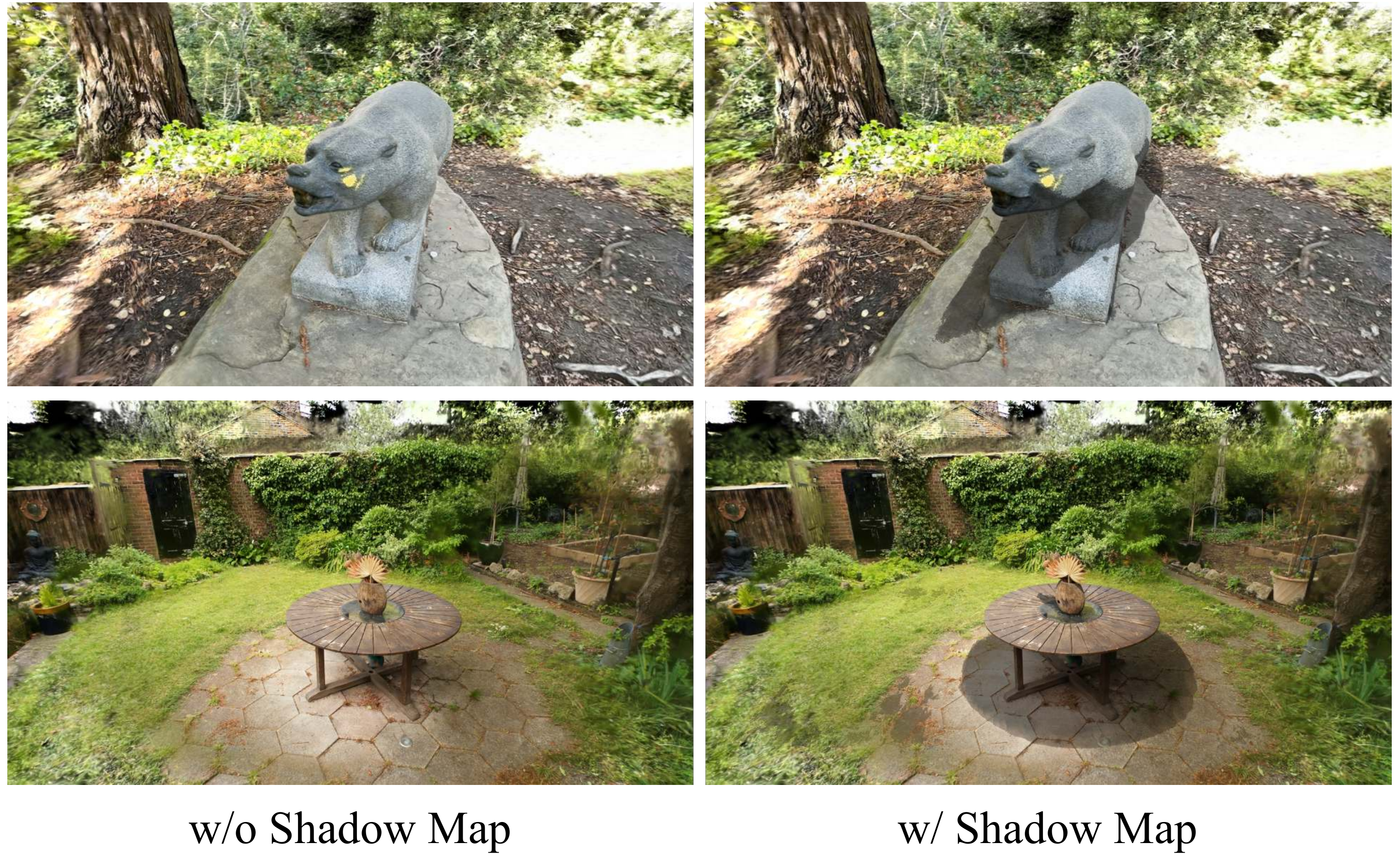}
  \caption{\textbf{Shadow Map Evaluation. } \ac{gs} traditionally models shadows as static surface textures. Our approach, employing a shadow map, generates dynamic shadows for a more immersive experience.}
  \label{fig:shadow_map}
\end{figure}

\paragraph{Shadow Map}\label{sec:ablation_shadow_map}
Furthermore, we take advantage of the shadow map to add extra time-dependent shadows into the \ac{gs} scene, as shown in \ref{fig:shadow_map}. Original \ac{gs} represents shadows as textures and thus fails to provide dynamic shadows when objects move. VR-GS allows users to choose the parameters, e.g. position and direction, of the light source to reproduce the lighting setting of the original scene, leading to a more realistic \ac{vr} environment. 

\section{Experiment}\label{sec:exp}

In this section, we benchmark our VR-GS system's simulation performance against other methods in GS and NeRF manipulation. Additionally, we showcase interactive demonstrations on VR devices. Our prototype, developed in Unity with a CUDA-implemented plugin, is tested using a Quest Pro Head-Mounted Display (HMD) and corresponding controllers, on an Intel Core i9-14900KF CPU with 32GB memory and an NVIDIA GeForce RTX 4090 GPU.

\subsection{Performance}
\begin{figure*} [ht]
 \includegraphics[width=1.0 \textwidth]{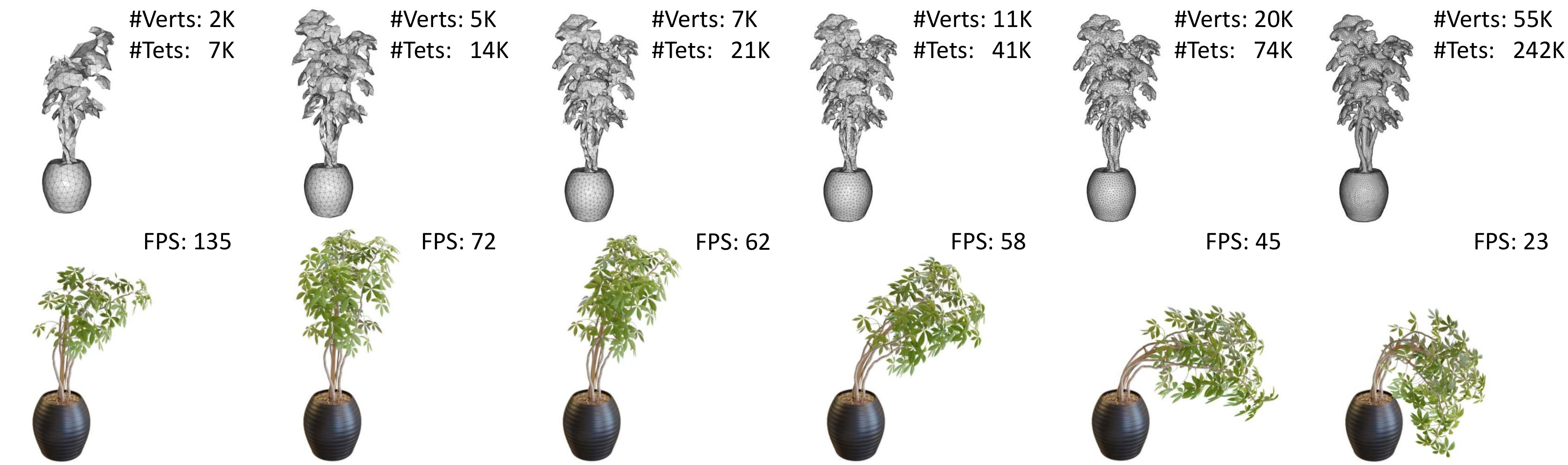}
  \centering
  \caption{\textbf{Trade-offs between Quality and Performance.} The top row displays cage meshes at varying resolutions, and the bottom row illustrates the corresponding simulation dynamics. Low-resolution meshes fail to capture fine dynamic details, whereas high-resolution meshes compromise real-time performance and can result in overly soft artifacts in the simulated object due to non-converging simulations. We employ mid-resolution meshes in practice to achieve an optimal balance between high frame rates and realistic physical dynamics.
  } 
  \label{fig:trade-offs}
\end{figure*}

VR-GS is designed for real-time, physics-aware interactions by integrating Gaussian kernels within a cage mesh for simulation. This cage mesh is derived from the VDB representation of Gaussians, with adjustable mesh resolution to influence quality. We first conduct experiments to explore the trade-offs between mesh quality and system performance. As shown in \Cref{fig:trade-offs}, using a coarse cage mesh facilitates higher frame rates but may lead to the loss of fine details. Conversely, a higher-resolution mesh inevitably increases the computational cost. Moreover, \ac{pbd} also requires much more iterations to achieve convergence for finer meshes. Without sufficient iterations, the simulated object may appear to be overly soft, yielding unrealistic dynamics. In practice, we constrain the number of mesh vertices to between 10K and 30K to maintain a balance between high frame rates and accurate physical dynamics.

\begin{figure}
  \includegraphics[width=1.0\linewidth]{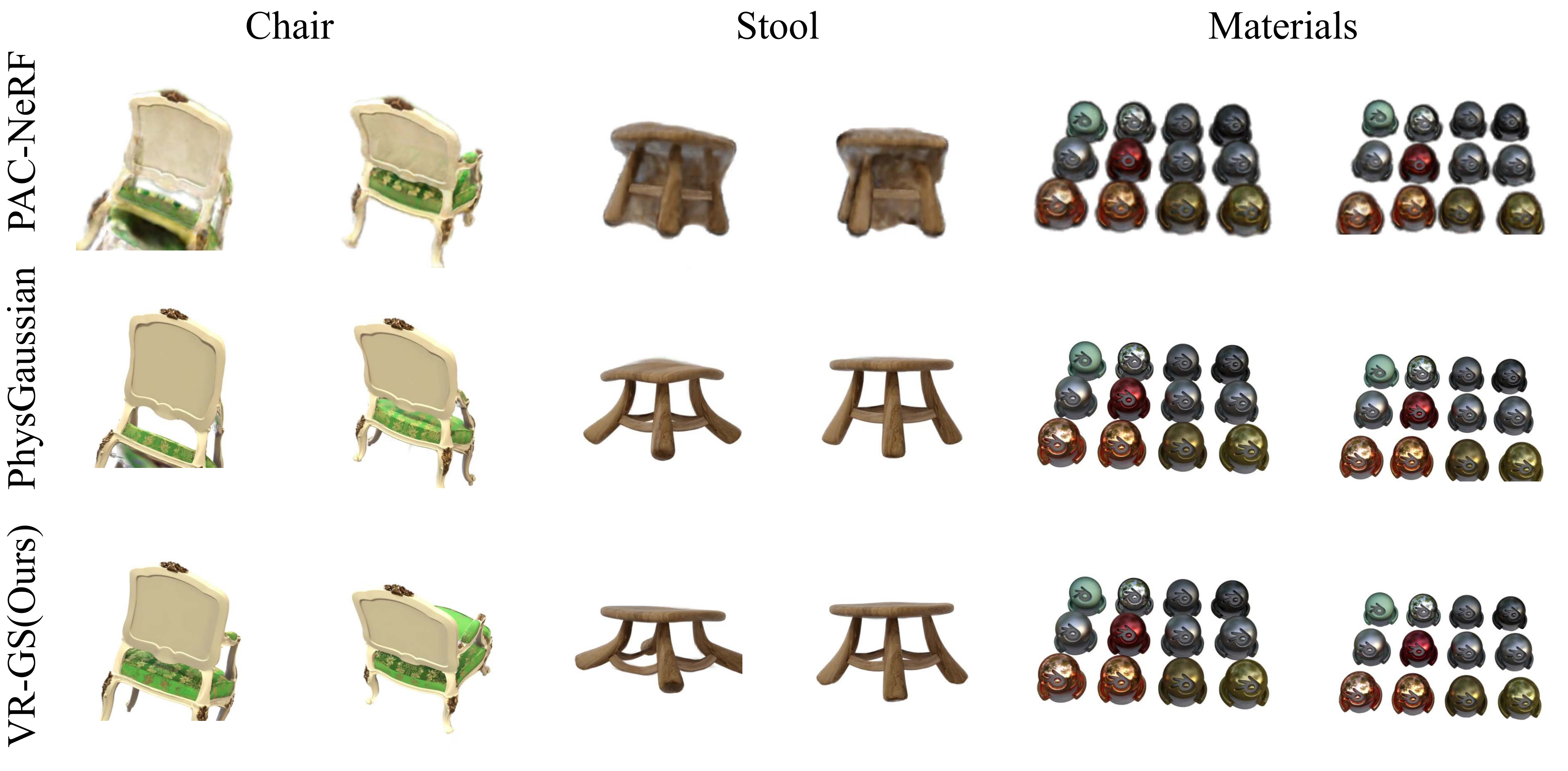}
  \caption{\textbf{Visual Quality Comparison.} Our method synthesizes competitive visual results compared to PhysGuassian \cite{xie2023physgaussian} and significantly outperforms PAC-NeRF \cite{li2023pac}.}
  \label{fig:comparison}
\end{figure}

\begin{wraptable}{r}{0.5\linewidth}
\hspace{-1.5em}
\resizebox{1.1\linewidth}{!}{
    \begin{tabular}{l|ccc}
    \hline
    Example & \begin{tabular}[c]{@{}l@{}}PAC- \\ NeRF \end{tabular}  & \begin{tabular}[c]{@{}l@{}}Phys- \\ Gaussian\end{tabular}   & \begin{tabular}[c]{@{}l@{}}VR-GS \\ \textbf{(Ours)} \end{tabular}  \\
    \hline
    Stool   & 0.750  & 0.112           & 0.017\\
    Chair   & 0.813    & 0.219           & 0.022 \\
    Materials     & 0.625    & 0.39           & 0.021\\
    \hline
    \end{tabular}
    }
\end{wraptable}
We then compare VR-GS against two state-of-the-art NeRF/GS physics-based manipulation methods: PAC-NeRF \cite{li2023pac} and PhysGaussian \cite{xie2023physgaussian}. PAC-NeRF primarily concentrates on estimating material parameters from multi-view videos in reconstructed \ac{nerf} scenes. Although it offers novel dynamic generation capabilities, the resulting visuals often fall short in rendering quality. In contrast, PhysGaussian, leveraging GS, produces superior photorealistic dynamics. However, the Material Point Method (MPM) used by both systems can limit real-time performance in complex scenes. For a fair comparison, we standardize the frame time ($\Delta t_{\text{frame}}=1/25$ sec) and the simulation step time ($\Delta t_{\text{step}}=0.0001$ sec) across all methods. For VR-GS, we set the XPBD iteration per substep to $1$, as the small $\Delta t_{\text{step}}$ sufficiently resolves the dynamics. As depicted in \Cref{fig:comparison} and the inset table, VR-GS not only matches the visual quality of PhysGaussian but also outperforms PAC-NeRF in clarity and realism. Crucially, our \ac{pbd}-based simulation framework allows for significantly faster simulations, making VR-GS ideal for real-time physics-aware interactions.

\subsection{VR Demos}\label{sec:demos}

\begin{figure*} [ht]
 \includegraphics[width=1.0\textwidth]{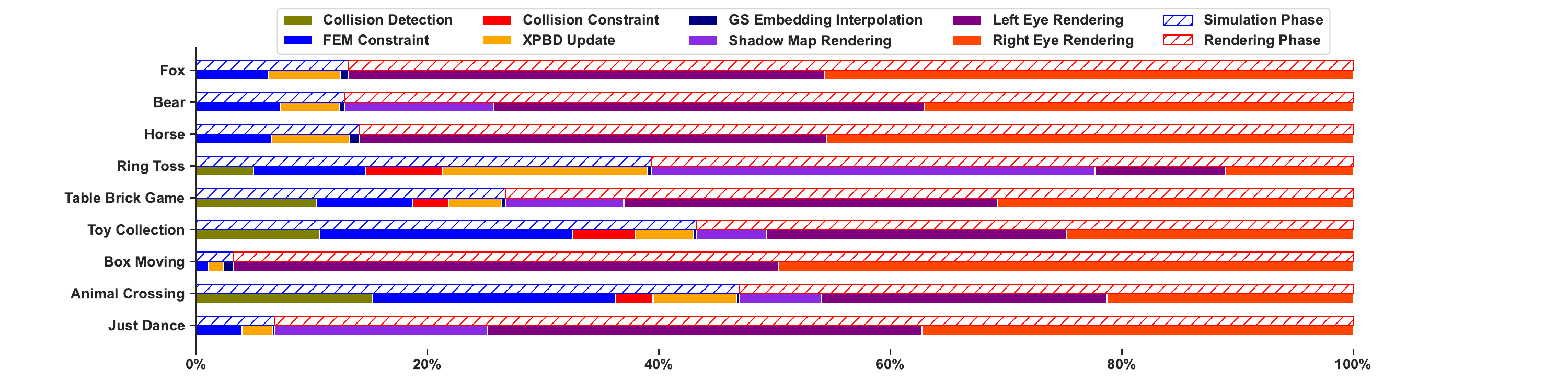}
  \centering
  \caption{\textbf{Timing Breakdown of Demos in \Cref{sec:demos}.}}
  \label{fig:timing_breakdown}
\end{figure*}

We then showcase VR-GS's capability to replicate real-world scenarios through several representative demos in VR. Detailed simulation setups and a timing breakdown for these demos are provided in \Cref{table:timing} and \Cref{fig:timing_breakdown}.

\paragraph{Fox, Bear, and Horse Manipulation} VR-GS empowers users to edit 3D Gaussian kernels intuitively and efficiently, utilizing our XPBD-based physics engine for dynamic manipulation. We demonstrate this through three examples as depicted in \Cref{fig:fox_bear_horse}: a fox, a bear, and a horse, reconstructed from the Instant-NGP dataset \cite{muller2022instant}, the Instruct-NeRF2NeRF \cite{haque2023instruct}, and the Tanks and Temples dataset \cite{knapitsch2017tanks}, respectively.

\paragraph{Ring Toss and Table Brick Game} 
In this example, we demonstrate the system's ability to seamlessly integrate new \emph{virtual objects} into existing scenes. We use a room scene reconstructed from real-world footage and virtual objects (rings and bricks modeled in Blender) to create ring tossing and table brick game (\Cref{fig:ring_toss}). The dining table and objects on it serve as collision boundaries for the elastic ring and rigid bricks \cite{deul2016position}.

\begin{table}[t]
    \centering
    \caption{\textbf{Parameters and Timings of Demos in \Cref{sec:demos}.}}
    \label{table:timing}
    \resizebox{0.9\linewidth}{!}{
    \begin{tabular}{l|rrccc}
    \hline
    Example & \#Gaussians  & \#Verts. &  \#Iter\textsuperscript{\small{*}} & CD\textsuperscript{\small{*}}  & FPS \\
    \hline
    (\Cref{fig:fox_bear_horse}) Fox & 304,875 & 28,205 & 1 & / & 161.2 \\
    (\Cref{fig:fox_bear_horse}) Bear & 2,525,891 & 19,472 & 5 & / & 76.9 \\
    (\Cref{fig:fox_bear_horse}) Horse & 1,295,223 & 10,611 &  5 & / & 115.4 \\
    (\Cref{fig:ring_toss}) Ring Toss & 1,456,209 & 9,002 & 10 & 10 & 37.7 \\
    (\Cref{fig:ring_toss}) Table Brick Game & 1,866,835 & 33,279 & 10 & 10 & 41.4 \\
    (\Cref{fig:toy_collection}) Toy Collection & 1,665,128 & 46,428 & 20 & 20 & 24.3 \\
    (\Cref{fig:inpainting}) Box Moving & 1,769,412 & 1,434 & 10 & / & 73.8 \\
    (\Cref{fig:teaser}) Animal Crossing & 1,312,670 & 36,386 & 10 & 5 & 34.7 \\
    (\Cref{fig:dance}) Just Dance & 1,059,054 & 13,936 & 50 & / & 33.9 \\
    \hline 
    \end{tabular}
    }
    \\ \small{\textsuperscript{*}\#Iter: XPBD iterations per substep; CD: collision detections per step.}
\end{table}

\paragraph{Toy Collection and Animal Crossing} We reconstructed a living room scene and a set of plush toys (\Cref{fig:toy_collection}) for users to interact with. The yellow spherical toy is generated using a text-to-3D generator LucidDreamer \cite{liang2023luciddreamer}, while the others are reconstructed. 
Additionally, four animal plush toys were placed on chairs, allowing users to manipulate and deform them freely (\Cref{fig:teaser}).

\begin{figure*} [ht]
 \includegraphics[width=1.0\textwidth]{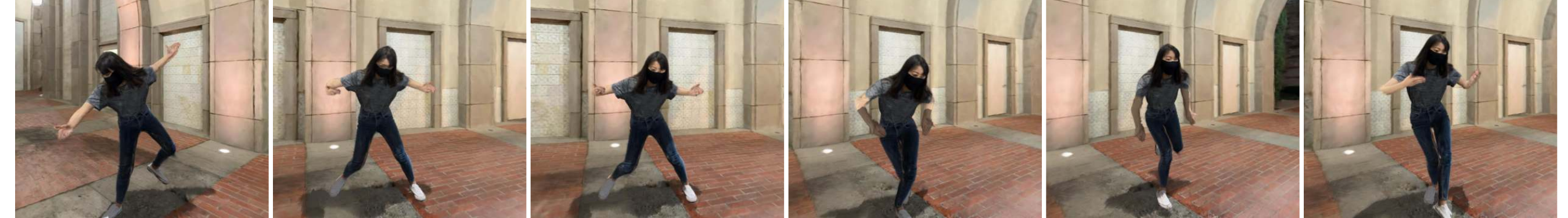}
  \centering
  \caption{\textbf{Just Dance.} VR-GS generates high-fidelity dance given a reconstructed human body.} 
  \label{fig:dance}
\end{figure*}

\paragraph{Just Dance} VR-GS also supports animation of reconstructed GS humans (\Cref{fig:dance}). Utilizing a reconstructed human character, we leverage Mixamo\footnote{https://www.mixamo.com/}'s auto-rig to generate motion sequences of its surface mesh. These sequences then serve as the boundary condition for the simulation, finally producing the GS human animation that can be viewed in our immersive system.

\subsection{Modeling Details}

We employ the \emph{Toy Collection} (\Cref{fig:toy_collection}) as a case study to demonstrate the time and labor required to create a fully interactive VR environment from scratch. The setup process is as follows:
\begin{enumerate}
\item The initial stage involves setting up the real-world scene. For the \emph{Toy Collection}, the indoor scene was captured directly as one scene, with all toys separately hung with strings as another scene. We completed it within 20 minutes.
\item Next, we capture multi-view images for Gaussian Splatting by recording videos of the two scenes with a consumer-level camera and converting them into image sequences. Camera intrinsic and extrinsic parameters are collected using COLMAP, with the entire process taking around 60 minutes, of which COLMAP occupies approximately 40 minutes.
\item We then proceed to image processing for segmentation and inpainting. Object classes are manually designated in the first video frame. Subsequent frames are automatically segmented using the method from \citet{cheng2023putting}, taking about 2 minutes. The inpainting of what's below the basket is fully automated \cite{suvorov2022resolution} and requires 5 minutes.
\item GS reconstruction is performed on the processed images, undergoing 30,000 training steps as specified by \citet{kerbl20233d} without additional treatment for the number of Gaussians, which takes about 30 minutes on a single RTX 4090.
\item After generating 3D GS, we execute internal filling, VDB reconstruction, and TetGen, all within 10 minutes.
\item Lastly, object world positions and physical parameters are specified and estimated in Unity, as detailed in Section \ref{sec:physparam}, taking 20 minutes.
\end{enumerate}
In summary, the total preparation time for the \emph{Toy Collection} is approximately 2 hours and 30 minutes. Manual tasks include scene staging, video recording, object class specification, and physical parameter selection. All other processes are automated. Once modeled, the scene is ready for a physics-aware interactive VR experience.

\section{User Study}

We conducted a user study including two tasks with 10 participants: 2 professionals (P1–P2) with 5 and 7 years of experience in 3D VFX software (including Houdini and Blender), and 8 novices (P3–P10), with P2 also having 2 years of VR experience. Our study used the hardware specified in the experimental section and included the Fox, Bear, House, Ring Toss, and Toy Collection demos. Participants received a 10-minute video tutorial on our system before completing two tasks. After task 2, they answered a questionnaire covering usability (System Usability Scale \cite{bangor2009determining}) and subjective feedback on individual and overall system features, as shown in \Cref{sec:user_res}. Additionally, we conducted a 20-minute semi-structured interview for more in-depth feedback.

\subsection{Tasks}
\begin{figure*}
  \includegraphics[width=1.0\linewidth]{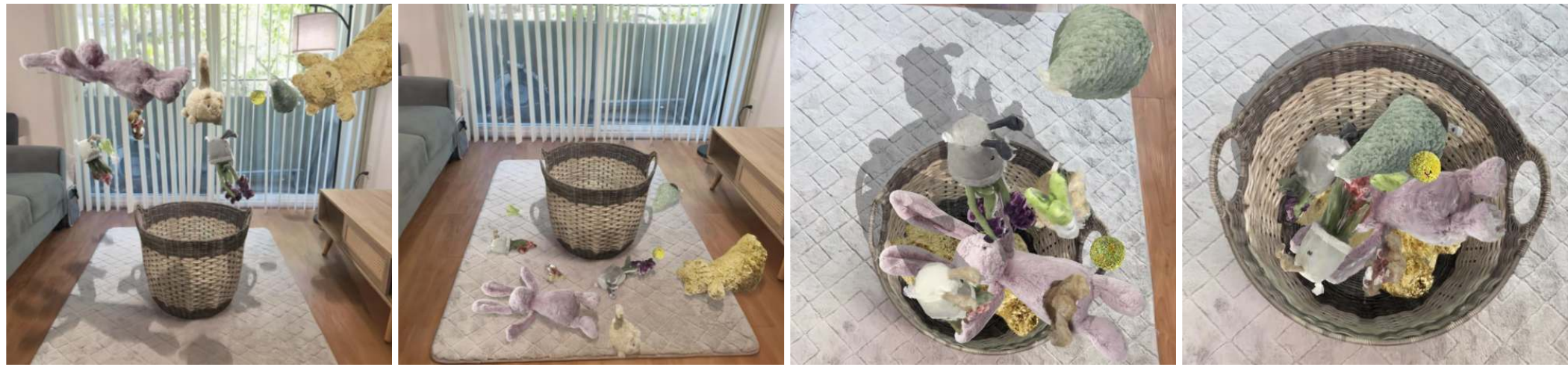}
  \caption{\textbf{Toy Collection.} We place all the toys on the ground first, and then move them into the basket.
  }
  \label{fig:toy_collection}
\end{figure*}

\paragraph{Task 1: Goal-directed (30 minutes).} The participants were required to play two VR games developed by our system: ring toss and toy collection (Figure \ref{fig:toy_collection}) in two different settings (S1: physics-based and S2: transform-based interaction). Each game had a 5-minute duration under each configuration. A 5-minute break was allowed between sessions. The arrangement of sessions and tasks used a Latin square design to counter potential learning effects. We requested users to rate the immersive and realistic experience in VR using a 7-point Likert scale.

\begin{figure*} [ht]
 \includegraphics[width=1.0\textwidth]{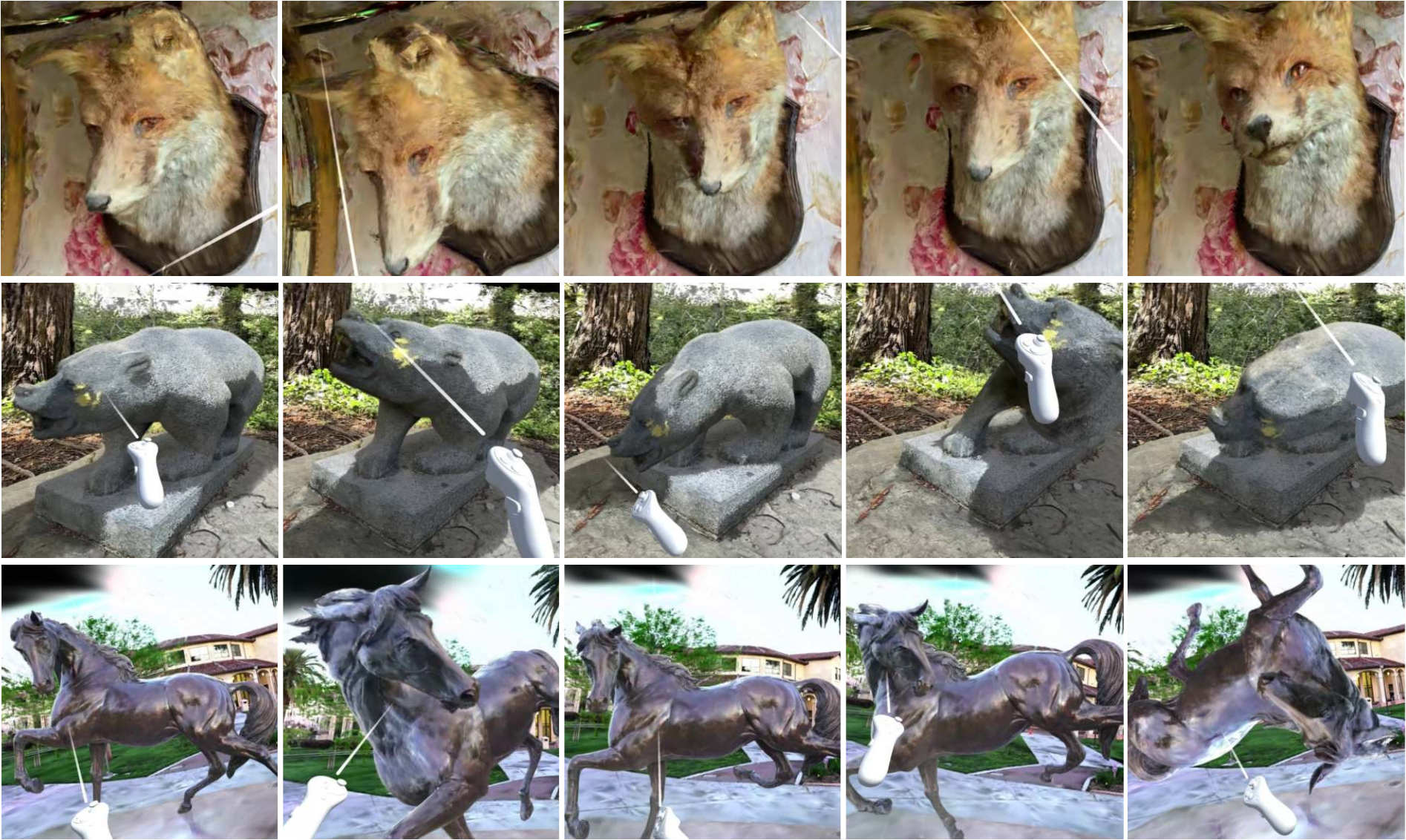}
  \centering
  \caption{\textbf{Fox, Bear, and Horse}. VR-GS allows users to manipulate 3D \ac{gs} at a real-time rate with physically plausible responses.} 
  \label{fig:fox_bear_horse}
\end{figure*}

\begin{figure*}
  \includegraphics[width=1.0\linewidth]{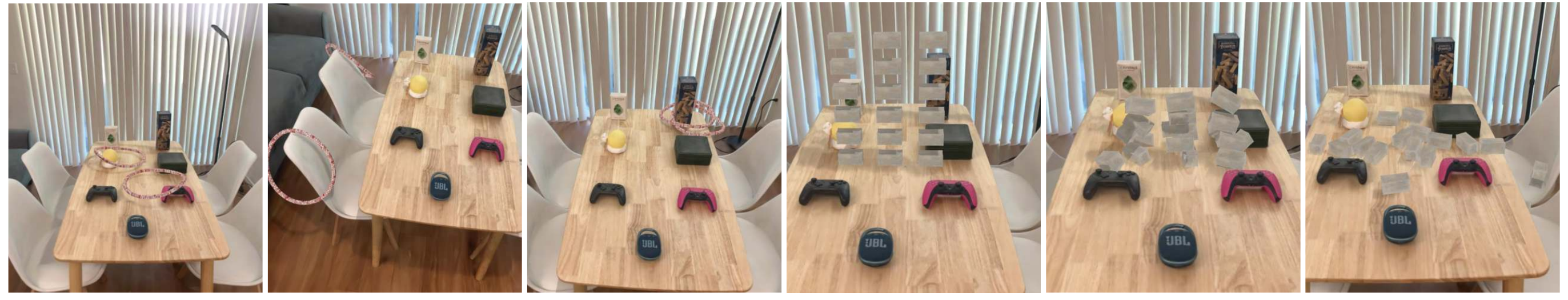}
  \caption{\textbf{Ring Toss and Table Brick Game.} Participants can engage in interactive ring toss and table brick breakout games simulated within authentic environments.}
  \label{fig:ring_toss}
\end{figure*}

\paragraph{Task 2: Open-ended (30 minutes).} The participants are asked to edit the scene and generate dynamics freely in aforementioned scenes (shown in Figure \ref{fig:fox_bear_horse}), Figure \ref{fig:ring_toss} and Figure \ref{fig:toy_collection}) with our prototype system, which includes geometry-based editing, physics-based editing, transform, rotation, duplicating, undoing, rescaling, etc. Task 2 is designed to evaluate the detailed potential factors impacting VR immersive experience.

\subsection{Results and Discussion}\label{sec:user_res} 

\begin{figure*} [ht]
 \includegraphics[width=1.0\textwidth]{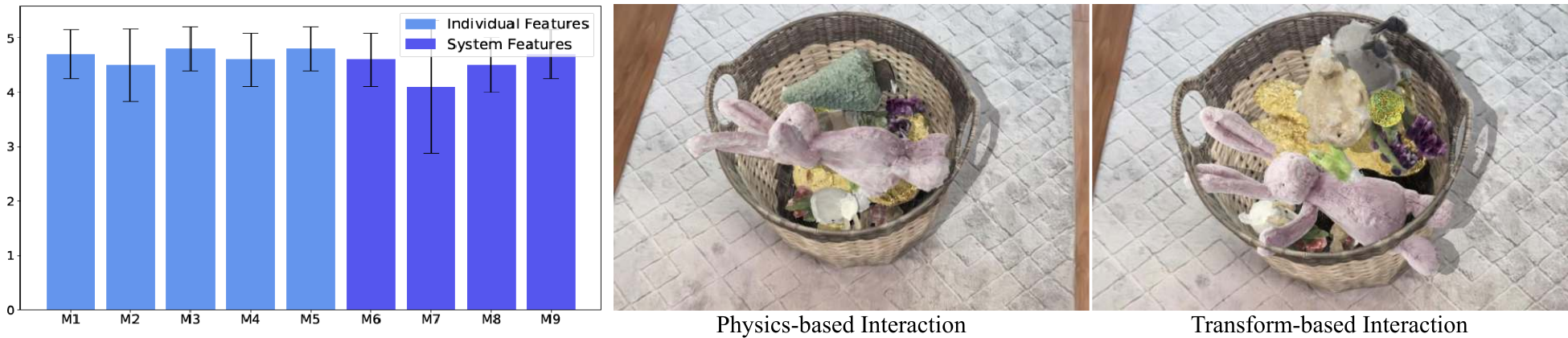}
  \centering
  \caption{\textbf{Study Results.} The left chart summarizes individual (M1-M5: object manipulation, scene inpainting, illumination, dynamics, physics placement) and system feedback (M6-M9: ease of use, latency, functionality, satisfaction). The right two figures show that physics-based interaction enhances immersion and realism in editing, whereas transform-based interaction yields less authentic outcomes, e.g. an undeformed and penetrated toys.} 

  \label{fig:studyres}
\end{figure*}

According to a paired t-test conducted on the score collected from task 1, it could be noticed that physics-based interaction significantly enhanced user immersion and realism compared to transform-based interaction (physics-based: 6.1 vs transform-based: 4.8 on average, \emph{p = .0227}). As shown in \Cref{fig:studyres}, our system has received overall positive opinions. Users spoke highly of the placement of virtual content. Transforming objects without physics rules resulted in floating objects, while in our system, all virtual content placements, such as putting toys onto a sofa, are followed by physics rules and are consistent with the user’s understanding. P7 commented, \textit{``I love this a lot. I have a dog. Petting the fox is really like what I did to my dog at home. I felt so real. Moreover, the placement of the toy is awesome. After they all fell to the basket, the basket even shook for a while.''} The VR-GS system's realistic lighting and physics-based dynamics create an immersive experience. As a professional user, P2 spoke highly of high-fidelity generative dynamics and illumination. \textit{``Those motions of virtual content are just like what I see in the physical world. I can't wait to take photos of my own house and then put them in VR with my video game character.''} Users rated the system highly on ease of use (4.6/5) and overall satisfaction (4.8/5). With a System Usability Scale (SUS) score of 83.5, VR-GS is classified as ``excellent'' according to \cite{bangor2009determining}. 

\section{Conclusion and Future Work}\label{sec:cfw}
We presented a physical dynamics-aware interactive Gaussian Splatting system for addressing challenges in editing real-time high-fidelity virtual content. By leveraging the advancements in Gaussian Splatting, VR-GS bridges the quality gap traditionally observed between machine-generated and manually created 3D content. Our system not only enhances the realism and immersion via physically-based dynamics, but also provides fine-grained interaction and manipulation controllability.

Although all the study participants appreciated the efficiency and effectiveness of our system, it still remains to be improved. Firstly, rendering high-fidelity Gaussian kernels in VR is computationally demanding. As a result, rendering generative dynamics in a large scene with 2K resolution might lead to potential latency issues in our system. Secondly, the physical parameters in our system are manually defined. Estimating parameters from videos like PAC-NeRF \cite{li2023pac}, or leveraging large-vision models, would be interesting directions to automate this process. In future work, we aim to incorporate a broader range of materials, such as fluid \cite{feng2024gaussian} and cloth, to enhance the system's capabilities. Furthermore, it is also interesting to explore how to utilize large multimodal models to assess the fidelity of the generated dynamics.

\begin{acks}
We thank the anonymous reviewers for their valuable feedback. We acknowledge support from NSF (2301040, 2008915, 2244651, 2008564, 2153851, 2023780), UC-MRPI, Sony, Amazon, and TRI.
\end{acks}

\bibliographystyle{ACM-Reference-Format}
\bibliography{vrgs}

\newpage

\end{document}